\documentclass[preprint,eqsecnum,aps,nofootinbib]{revtex4}
\usepackage{amsfonts,amsmath,amssymb,amsthm}
\usepackage{latexsym}
\usepackage{bbm,bm}
\usepackage{graphicx}

\newcommand{\ket}[1]{\lvert #1 \rangle}
\newcommand{\bra}[1]{\langle #1 \lvert}
\newcommand{\beq}{\begin{equation}}
\newcommand{\eeq}{\end{equation}}
\newcommand{\beqs}{\begin{eqnarray}}
\newcommand{\eeqs}{\end{eqnarray}}
\begin{document}

\title{Scrambling and Quantum Teleportation}

\author{MuSeong Kim$^1$, Mi-Ra Hwang$^1$, Eylee Jung$^1$, and DaeKil Park$^{1,2}$\footnote{corresponding author, dkpark@kyungnam.ac.kr} }

\affiliation{$^1$Department of Electronic Engineering, Kyungnam University, Changwon,
                 631-701, Korea    \\
             $^2$Department of Physics, Kyungnam University, Changwon,
                  631-701, Korea }

%\author{DaeKil Park$^{1,2}$\footnote{dkpark@kyungnam.ac.kr} }
%
%\affiliation{$^1$Department of Electronic Engineering, Kyungnam University, Changwon
%                631-701, Korea    \\
%           $^2$Department of Physics, Kyungnam University, Changwon
%                 631-701, Korea    
%                      }

\begin{abstract}
Scrambling is a concept introduced from information loss problem arising in black hole. In this paper we discuss the effect of scrambling
from a perspective of pure quantum information theory. We introduce $7$-qubit quantum circuit for a quantum teleportation. It is shown that the teleportation can be 
perfect if a maximal scrambling unitary is used. From this fact we conjecture that ``the quantity of scrambling is proportional to the fidelity of teleportation''. In order to 
confirm the conjecture we introduce $\theta$-dependent partially scrambling unitary, which reduces to no scrambling and maximal scrambling at $\theta = 0$ and $\theta = \pi / 2$, respectively.
Then, we compute the average fidelity analytically, and numerically by making use of qiskit (version $0.36.2$)  and $7$-qibit real quantum computer ibm$\_$oslo.
Finally, we conclude that our conjecture can be true or false depending on the choice of qubits for Bell measurement.

\end{abstract}
\maketitle

\section{Introduction}
Nowadays, quantum information theories (QIT)\cite{text} is one of the subjects, which attract much attention recently. This seems to be mainly due to the rapid development of quantum technology such as realization of quantum cryptography\cite{cryptography2,white} and quantum computer\cite{qcreview,ibm}.
In QIT quantum entanglement\cite{schrodinger-35,text,horodecki09} plays an important role as a physical resource in the various  types of quantum information processing (QIP).
It is used in  many QIP such as  in quantum teleportation\cite{teleportation,Luo2019},
superdense coding\cite{superdense}, quantum cloning\cite{clon}, quantum cryptography\cite{cryptography,cryptography2}, quantum
metrology\cite{metro17}, and quantum computers\cite{qcreview,computer,supremacy-1}. In particular, quantum computing attracted a lot of  attention recently after IBM and Google independently realized quantum computers.
It is debatable whether  ``quantum supremacy'' is achieved or not in the quantum computation. 

%As IC (integrated circuit) becomes smaller and smaller in modern classical technology, the effect of quantum mechanics becomes prominent more and more. As a result, quantum technology (technology 
%based on quantum mechanics and quantum information theories\cite{text}) becomes important more and more recently. The representative constructed by quantum technology
%is a quantum computer\cite{supremacy-1}, which was realized recently by making use of superconducting qubits. 
%In quantum information processing quantum entanglement\cite{text,schrodinger-35,horodecki09} plays an important role as a physical resource. 
%It is used in various quantum information processing, such as  quantum teleportation\cite{teleportation,Luo2019},
%superdense coding\cite{superdense}, quantum cloning\cite{clon}, quantum cryptography\cite{cryptography,cryptography2}, quantum
%metrology\cite{metro17}, and quantum computer\cite{supremacy-1,qcreview,computer}. 
%Furthermore, with many researchers trying to realize such quantum information processing in the laboratory for the last few decades, quantum cryptography and quantum computer seem to approaching the commercial level\cite{white,ibm}.

Quantum Gravity (QG) is a field of physics that seeks to describe gravity according to the principles of quantum mechanics. 
Experimental access to QG, however, is challenging at present since it requires the ability to measure miniscule physical effects.
Recent rapid development of quantum computer, however, may allow different possibility to test QG indirectly. Using quantum simulators and quantum computers we may be able to probe QG
in the laboratory\cite{garcia17,ippei17,franz18,adam19}.

Already there were several papers investigating QG toward this direction. In Ref.\cite{enrico22} matrix quantum mechanics is simulated by adopting the quantum-classical hybrid algorithm called VQE\cite{VQE-1}. 
In particular, the authors of Ref.\cite{enrico22} computed the low-energy spectra of bosonic and supersymmetric matrix models and compare them to the results of Monte Carlo simulations. 
In Ref.\cite{kelvin18} quantum teleportation with scrambling unitary was implemented on a fully-connected trapped-ion quantum computer\cite{debnath16}. 
This is based on the Hayden-Preskill protocol\cite{hayden07,beni17,beni18}. The intuition behind their approach is to reinterpret the black hole's information loss problem via the quantum teleportation.
In Ref.\cite{illya22} wormhole-inspired teleportation was simulated by making use of Quantinuum's trapped-ion System Model H1-1 and five IBM superconducting quantum processing units.
This is indirect approach to verify the ER=EPR conjecture\cite{erepr1,erepr2}, which assumes that the quantum channel generated by entangled quantum state is nothing but the wormhole.
It was shown that the teleportation signals reach $80 \%$ of theoretical predictions. 

%%%%%%%%%%%%%%%%%%%%%%%%%%%%%%%%%%%%%%%%%%%%%%%%%%%%%%%%%
\begin{figure}[ht!]
\begin{center}
\includegraphics[height=5.0cm]{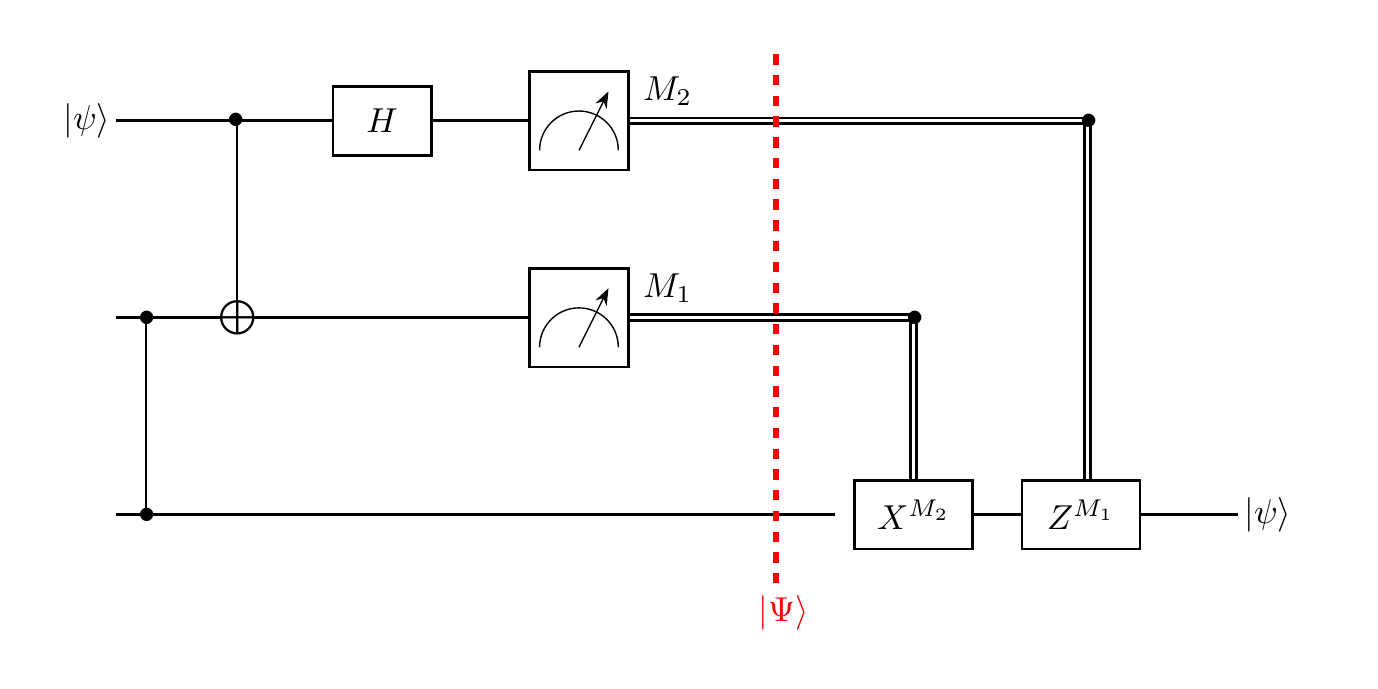} 
\includegraphics[height=6.0cm]{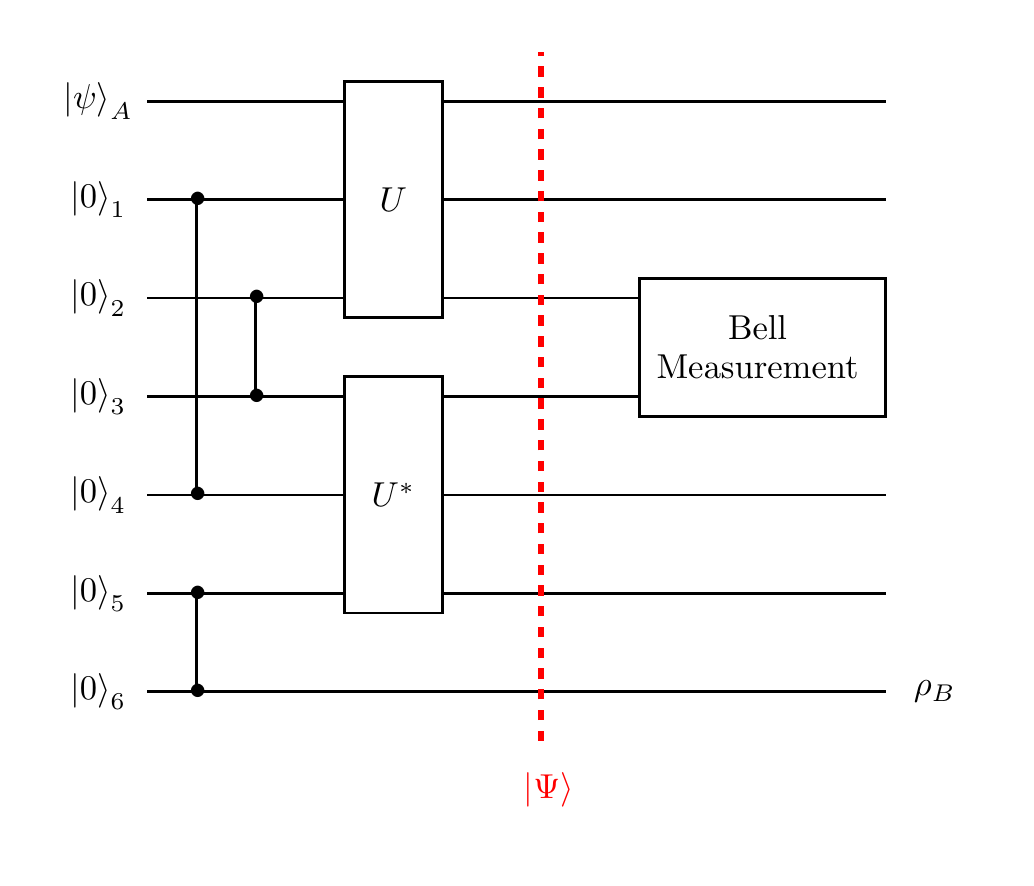}

\caption[fig1]{(Color online) (a) Quantum circuit for usual quantum teleportation.
(b) Quantum circuit for quantum teleportation with scrambling unitary. 
If $U$ is maximally scrambling unitary, perfect quantum teleportation might be possible. 
In both figures the vertical lines correspond to the initial maximally entangled state $\ket{\beta_0} = \frac{1}{\sqrt{2}} (\ket{00} + \ket{11})$. }
\end{center}
\end{figure}
%%%%%%%%%%%%%%%%%%%%%%%%%%%%%%%%%%%%%%%%%%%%%%%%%%%%%%%%%%%

In this paper we study the teleportation scheme with a scrambling unitary from a viewpoint of pure QIT. The scrambling\cite{beni17,beni18,scramble1,scramble2,scramble3} is a concept introduced 
from information loss problem\cite{hawking1,page1} in black hole physics. Although there is more rigorous definition\cite{scramble1,scramble2}, roughly speaking, ``scrambling'' means the delocalization of quantum information. In other words, when the quantum information of 
the subsystem is completely mixed with remaining systems, we use the terminology ``scrambling''\footnote{In the information loss problem the scrambling time is important to check the validity of quantum mechanics.} .

The quantum circuit for this scheme is different from usual quantum teleportation as shown in Fig, 1.
Fig. 1a is well-known $3$-qubit quantum circuit for usual quantum teleportation. Alice has first two qubits and Bob has last one. The vertical line means the maximally entangled state $\ket{\beta_0} = \frac{1}{\sqrt{2}} (\ket{00} + \ket{11})$.
The task is to teleport the unknown state $\ket{\psi} = \alpha \ket{0} + \beta \ket{1}$ to Bob. It is easy to show that the quantum state $\ket{\Psi}$ in Fig. 1a is 
\begin{equation}
\label{usualqt}
\ket{\Psi} = \frac{1}{2} \bigg[ \ket{00} (\alpha \ket{0} + \beta \ket{1}) + \ket{01} (\alpha \ket{1} + \beta \ket{0}) + \ket{10} (\alpha \ket{0} - \beta \ket{1}) + \ket{11} (\alpha \ket{1} - \beta \ket{0}) \bigg].
\end{equation}
Thus, the task is completed by applying $X$ and/or $Z$ to Bob's qubit appropriately, where $X$, $Y$, and $Z$ are the Pauli operators. 

Fig. 1b is a $7$-qubit quantum circuit for  teleportation with scrambling. 
First qubit, {\it i.e.} $0^{th}$-qubit, in Fig. 1b represents the Alice's secret qubit. In this paper we assume $\ket{\psi}_A = \alpha \ket{0} + \beta \ket{1}$ with $|\alpha|^2 + |\beta|^2 = 1$. 
The  $1^{th}$- and $2^{th}$-qubits are Charlie's  qubits. Thus, unitary operator $U$ scrambles the quantum information of Alice's and Charlie's qubits.
The  $3^{th}$- and $4^{th}$-qubits denote Daniel's qubits. Finally, the  $5^{th}$- and $6^{th}$-qubits are Bob's ancillary qubits\footnote{If Charlie' and Daniel's qubits are replaced with black hole's and Bob's quantum memory qubits respectively, this circuit can be used to explore the information loss problem.}.
The vertical lines in Fig. 1b  means $\ket{\beta_0}$ too. Of course, $U^*$ is a complex conjugate of unitary $U$. 
Here, we assume that Daniel can access to all parties. Therefore, Daniel can select the unitary operator $U$ and quantum measurement freely. 
Then, the question is as follows:  is it possible to teleport Alice's qubit $\ket{\psi}_A$ to Bob's $6^{th}$ qubit if Daniel selects $U$ and quantum measurement appropriately?

It was suggested\cite{kelvin18,hayden07,beni17,beni18} that if $U$ is chosen as maximally scrambling unitary, perfect teleportation might be possible if Daniel chooses a quantum measurement appropriately and notifies the outcome to Bob through a classical channel. 
If this is right, one can guess that if $U$ is partially scrambling unitary, the fidelity of teleportation is lowered from one even though Daniel performs the optimal quantum measurement.
This means that  the quantity of scrambling of $U$ is probably proportional to the fidelity for the teleportation. 
The purpose of the paper is to examine this conjecture. 
In order to explore this problem we introduce $U(\theta)$, where $\theta = 0$ and $\theta = \pi / 2$ correspond to the no scrambling and maximally scrambling.
Since there is no measure which quantify the scrambling, we cannot say how much quantum information is scrambled by $U(\theta)$. But from the parametrization in $\theta$, we guess that the quantity of scrambling of $U(\theta)$ is 
proportional to $\theta$. Then, we will compute the  $\theta$-dependence of the fidelities between $\ket{\psi}_A$ and $\rho_B$ analytically. In order to examine the noise effect we also compute the fidelities numerically by making use of 
qiskit (version $0.36.2$) and $7$-qubit real quantum computer ibm$\_$oslo. From the analytical and numerical results we conclude that our conjecture ``the quantity of scrambling is proportional to the fidelity of teleportation'' can be true or false depending on the Daniel's choice of qubits for Bell measurement.

The paper is organized as follows. In next section we examine the quantum teleportation with maximally scrambling $U$. If Daniel chooses Bell measurement in one of $\{2,3\}$, $\{1,4\}$ or $\{0,5\}$ qubits and notifies the outcomes to Bob, 
it is shown that the perfect teleportation is possible. In section III we examine the teleportation again with  $U(\theta)$, which is no scrambling at $\theta = 0$ and maximally scrambling at $\theta = \pi / 2$. 
If Daniel takes  Bell measurement of either  $\{2,3\}$ or $\{1,4\}$ qubits, it is shown that the fidelities are the exactly the same. The $\theta$-dependence of average fidelity is monotonically increasing function with respect to $\theta$, which
supports the conjecture. If, however, Daniel takes Bell measurement of $\{0,5\}$ qubits, it is shown that the average fidelity is not monotonic.
In section IV the numerical calculation for the fidelities is discussed. 
Comparing the analytically computed fidelities with the numerical ones, it is shown that qiskit and ibm$\_$oslo yields errors less that $1\%$. 
Therefore, the effect of noise is negligible in the calculation of fidelities. 
Thus, if we need to discuss a similar issue in the future with large number of qubits, we can adopt the numerical approach without producing much error.
In section V a brief conclusion is given. 
In appendix A the partial scrambling property of $U(\theta)$ is more clearly verified.  The numerical results are summarized in appendix B and appendix C.

\section{Quantum Teleportation with maximally scrambling unitary}

%%%%%%%%%%%%%%%%%%%%%%%%%%%%%%%%%%%%%%%%%%%%%%%%%%%%%%%%%
\begin{figure}[ht!]
\begin{center}
\includegraphics[height=3.5cm]{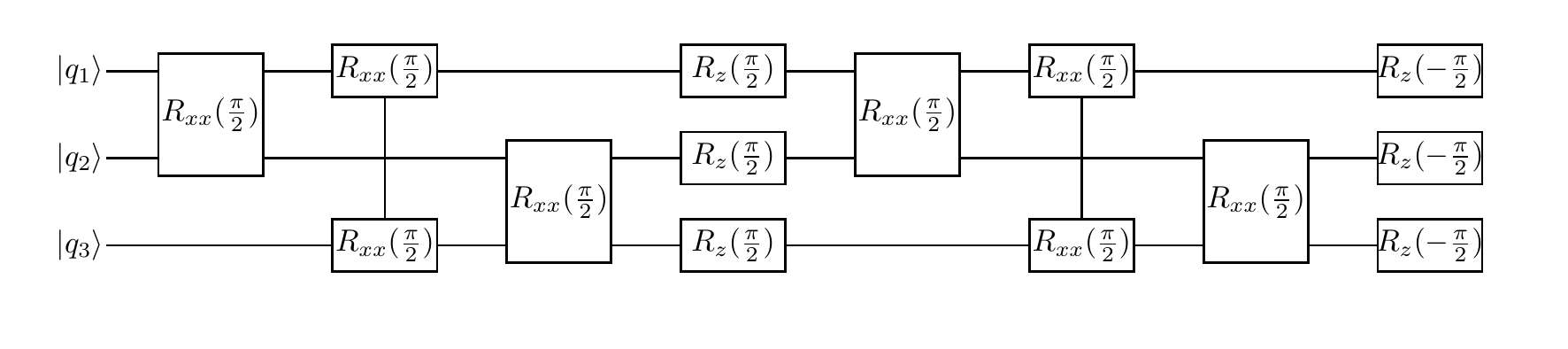} 

\caption[fig2]{(Color online) Quantum circuit for implementing the unitary $U$ in Eq.  (\ref{scram2-1}).
 }
\end{center}
\end{figure}
%%%%%%%%%%%%%%%%%%%%%%%%%%%%%%%%%%%%%%%%%%%%%%%%%%%%%%%%%%%

In this section we choose $U$ in a form:
\begin{eqnarray}
\label{scram2-1}
U = \frac{1}{2}   \left(       \begin{array}{cccccccc}
                                  -1 & 0 & 0 & -1 & 0 & -1 & -1 & 0    \\
                                  0 & 1 & -1 & 0 & -1 & 0 & 0 & 1       \\
                                  0 & -1 & 1 & 0 & -1 & 0 & 0 & 1       \\
                                  1 & 0 & 0 & 1 & 0 & -1 & -1 & 0       \\
                                  0 & -1 & -1 & 0 & 1 & 0 & 0 & 1       \\
                                  1 & 0 & 0 & -1 & 0 & 1 & -1 & 0       \\
                                  1 & 0 & 0 & -1 & 0 & -1 & 1 & 0                               \\
                                  0 & -1 & -1 & 0 & -1 & 0 & 0 & -1
                                          \end{array}                                                     \right).
\end{eqnarray}
This unitary operator can be experimentally implemented up to the global phase by a quantum circuit in Fig. 2.
It is straightforward to show 
\begin{eqnarray}
\label{scram2-2}
&&    U^{\dagger} (X \otimes I \otimes I) U = - X \otimes Z \otimes Z   \hspace{1.0cm}  U^{\dagger} (I \otimes X \otimes I) U = - Z \otimes X \otimes Z        \\        \nonumber
&&    U^{\dagger} (I \otimes I \otimes X) U = - Z \otimes Z \otimes X   \hspace{1.0cm}  U^{\dagger} (Y \otimes I \otimes I) U = - Y \otimes X \otimes X        \\        \nonumber
&&    U^{\dagger} (I \otimes Y \otimes I) U = - X \otimes Y \otimes X   \hspace{1.0cm}  U^{\dagger} (I \otimes I \otimes Y) U = - X \otimes X \otimes Y        \\        \nonumber
&&    U^{\dagger} (Z \otimes I \otimes I) U = - Z \otimes Y \otimes Y   \hspace{1.0cm}  U^{\dagger} (I \otimes Z \otimes I) U = - Y \otimes Z \otimes Y        \\        \nonumber
&&    \hspace{4.0cm}    U^{\dagger} (I \otimes I \otimes Z) U = - Y \otimes Y \otimes Z    
\end{eqnarray}
where $X$, $Y$, $Z$ and $I$ are the three Pauli operators and the Identity operator. 
Eq. (\ref{scram2-2}) verifies the maximal scrambling property of $U$ by showing that it delocalizes  all singlet-qubit into three-qubit operators. 
One can show that the quantum state $\ket{\Psi}$ in Fig. 1b is 
\begin{equation}
\label{qt2-1}
\ket{\Psi} = \frac{1}{2} \bigg[ b_{000} (\alpha \ket{0} + \beta \ket{1}) - b_{211} (\alpha \ket{0} - \beta \ket{1}) + b_{133} (\beta \ket{0} + \alpha \ket{1}) - b_{322}  (-\beta \ket{0} + \alpha \ket{1}) \bigg]
\end{equation}
up to global phase. 
In Eq. (\ref{qt2-1}) $b_{ijk} = \ket{\beta_i}_{05} \ket{\beta_j}_{14} \ket{\beta_k}_{23}$, where 
\begin{eqnarray}
\label{qt2-2}
&& \ket{\beta_0} = \frac{1}{\sqrt{2}} \left( \ket{00} + \ket{11} \right)   \hspace{1.0cm}  \ket{\beta_1} = \frac{1}{\sqrt{2}} \left( \ket{01} + \ket{10} \right)             \\   \nonumber
&&  \ket{\beta_2} = \frac{1}{\sqrt{2}} \left( \ket{00} - \ket{11} \right)  \hspace{1.0cm}  \ket{\beta_3} = \frac{1}{\sqrt{2}} \left( \ket{01} - \ket{10} \right).
\end{eqnarray}

%%%%%%%%%%%%%%%%%%%%%%%%%%%%%%%%%%%%%%%%%%%%%%%%%%%%%%%%%
\begin{figure}[ht!]
\begin{center}
\includegraphics[height=5.0cm]{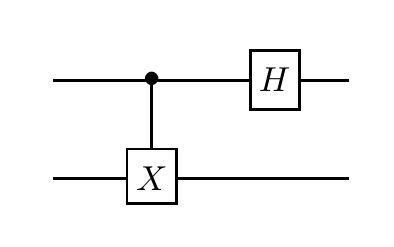} 

\caption[fig3]{(Color online)  Quantum circuit for Bell measurement, which generates Eq. (\ref{bell2-1}). 
 }
\end{center}
\end{figure}
%%%%%%%%%%%%%%%%%%%%%%%%%%%%%%%%%%%%%%%%%%%%%%%%%%%%%%%%%%%

From Eq. (\ref{qt2-1}) it is easy to show that the teleportation process is completed if Daniel performs a Bell measurement  in one of $\{2,3\}$, $\{1,4\}$, or $\{0,5\}$ qubits and notifies the measurement outcomes to Bob.
The Bell measurement can be easily implemented by using a quantum circuit of Fig. 3. 
This circuit transforms the Bell states into the computation basis as 
\begin{equation}
\label{bell2-1}
\ket{\beta_0} \rightarrow \ket{00}    \hspace{1.0cm} \ket{\beta_1} \rightarrow \ket{01}    \hspace{1.0cm} \ket{\beta_2} \rightarrow \ket{10}    \hspace{1.0cm} \ket{\beta_3} \rightarrow \ket{11}.
\end{equation}

Let us assume that Daniel chooses $\{2,3\}$ or $\{1,4\}$ qubits as a Bell measurement. If the measurement outcomes are 
$(0,0)$, $(0,1)$, $(1.0)$, or $(1,1)$, Bob's $6^{th}$-qubit state can be $\ket{\psi}_A$ if Bob operates $I$, $Z$, $ZX$, or $X$ to his qubit.  
If Daniel takes $\{0,5\}$ qubits, Bob's state also can be  $\ket{\psi}_A$ by operating $I$, $X$, $Z$, or $ZX$ to his qubit. 
Therefore, the maximal scrambling unitary $U$ given in Eq. (\ref{scram2-1}) really allows the perfect teleportation.

\section{Quantum Teleportation with partial scrambling unitary}

%%%%%%%%%%%%%%%%%%%%%%%%%%%%%%%%%%%%%%%%%%%%%%%%%%%%%%%%%
\begin{figure}[ht!]
\begin{center}
\includegraphics[height=3.5cm]{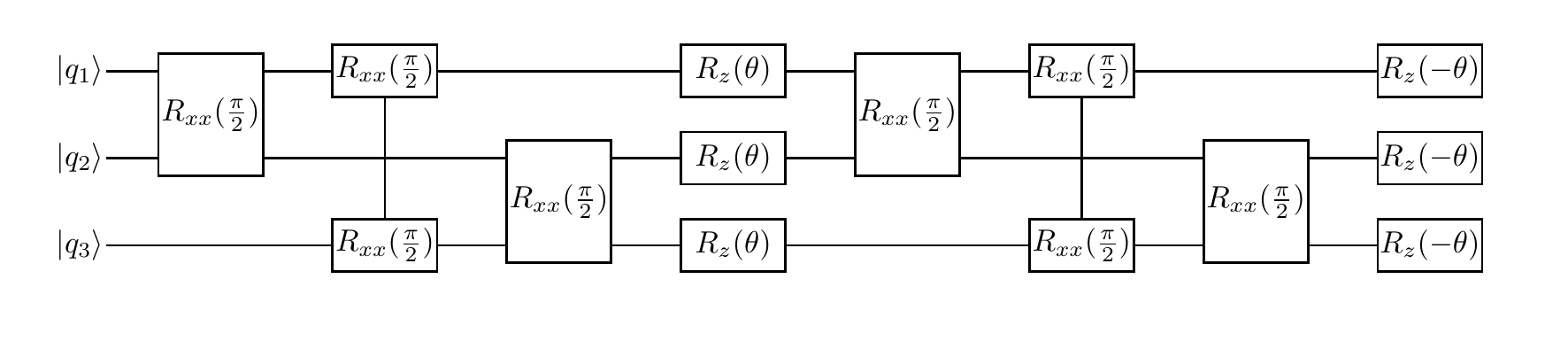} 

\caption[fig4]{(Color online) Quantum circuit for implementing the unitary in Eq. (\ref{scram3-1}).}
\end{center}
\end{figure}
%%%%%%%%%%%%%%%%%%%%%%%%%%%%%%%%%%%%%%%%%%%%%%%%%%%%%%%%%%%

In the previous section we showed that perfect quantum teleportation is possible if the maximal scrambling unitary (\ref{scram2-1}) is used. In order to understand the role of scrambling property in the teleportation process more clearly, we consider in this section the teleportation with 
partial scrambling unitary. For this purpose we choose $U$ as a $\theta$-dependent unitary in the form:
\begin{eqnarray}
\label{scram3-1}
U = \frac{1}{4} \left(                                       \begin{array}{cccccccc}
                               \mu_{2,+} & 0 & 0 & -\mu_{1,+} & 0 & -\mu_{1,+} & -\mu_{1,+} & 0                 \\
                               0 & \mu_{3,+} & -\mu_{1,+} & 0 & -\mu_{1,+} & 0 & 0 & \mu_{1,+}                    \\
                               0 & -\mu_{1,+} & \mu_{3,+} & 0 & -\mu_{1,+} & 0 & 0 &  \mu_{1,+}                  \\
                               \mu_{1,-} & 0 & 0 & \mu_{3,-} & 0 & -\mu_{1,-} & -\mu_{1,-} & 0                      \\
                               0 & -\mu_{1,+} & -\mu_{1,+} & 0 & \mu_{3,+} & 0 & 0 & \mu_{1,+}                   \\
                               \mu_{1,-} & 0 & 0 & -\mu_{1,-} & 0 & \mu_{3,-} & -\mu_{1,-} & 0                   \\
                                \mu_{1,-} & 0 & 0 & -\mu_{1,-} & 0 & -\mu_{1,-} & \mu_{3,-} & 0                  \\
                                0 & -\mu_{1,-} & -\mu_{1,-} & 0 & -\mu_{1,-} & 0 & 0 & \mu_{2,-}
                                                                       \end{array}                                                  \right)
\end{eqnarray}
where
\begin{equation}
\label{scram3-2}
\mu_{1,\pm} = 1 - e^{\pm 2 i \theta}   \hspace{1.0cm}  \mu_{2,\pm} = 1 + 3 e^{\pm 2 i \theta}  \hspace{1.0cm}  \mu_{3,\pm} = 3 + e^{\pm 2 i \theta}.
\end{equation}
When $\theta=0$, it reduces to the identity, which has no scrambling property. When $\theta = \pi / 2$, it reduces to Eq. (\ref{scram2-1}), which has a maximal scrambling property. 
When $0 < \theta < \pi / 2$, this is a partially scrambling unitary. This fact can be explicitly verified by examining how the maximal scrambling property (\ref{scram2-2}) is modified if $U$ is replaced by Eq. (\ref{scram3-1}). This is summarized in appendix A. 
The unitary (\ref{scram3-1}) can be implemented up to the global phase by a quantum circuit in Fig. 4.
Then, one can show that the quantum state $\ket{\Psi}$ in Fig. 1b can be written in a form:
\begin{eqnarray}
\label{qt3-1}
&&\ket{\Psi} = \frac{1}{8} \Bigg[ 4 b_{000} (\alpha \ket{0} + \beta \ket{1})                                           \\    \nonumber
&&\hspace{2.0cm}  + \bigg\{ \sin^2 2 \theta (b_{112} + b_{121} + b_{233} + b_{323} + b_{332}) + 4 \cos^2 \theta b_{200}      \\    \nonumber
&& \hspace{2.7cm}     - 4 \sin^4 \theta b_{211}   + 4 i \sin^3 \theta \cos \theta (b_{123} + b_{132} + b_{213} + b_{231})                                                                        \\     \nonumber
&& \hspace{3.0cm}   - 4 i \sin \theta \cos^3 \theta (b_{312} + b_{321}) \bigg\} (\alpha \ket{0} - \beta \ket{1})                                                                              \\    \nonumber
&&\hspace{2.0cm} + \bigg\{ 4 \cos^2 \theta b_{100} - \sin^2 2 \theta (b_{111} + b_{313} + b_{331}) + 4 i \sin \theta \cos^3 \theta b_{311}    + 4 \sin^4 \theta b_{133}       \\    \nonumber
&& \hspace{2.7cm} -4 i \sin^3 \theta \cos \theta (b_{113} + b_{131} + b_{333}) + 2 i \sin 2\theta b_{300} \bigg\} (\beta \ket{0} + \alpha \ket{1})                                     \\    \nonumber
&&\hspace{2.0cm} + \bigg\{ 4 \cos^4 \theta b_{300} + 4 i \sin \theta \cos^3 \theta (b_{100} - b_{212} - b_{221})                                                                                        \\    \nonumber
&& \hspace{2.7cm} - 4 i \sin^3 \theta \cos \theta (b_{001} + b_{010} - b_{122}) - 4 \sin^4 \theta b_{322}                                                                                                      \\    \nonumber
&& \hspace{2.7cm} -\sin^2 2 \theta (b_{003} + b_{030} + b_{113} + b_{131} - b_{223} - b_{232 } + b_{311} - b_{333})                                                                                          \\    \nonumber
&& \hspace{4.0cm} + i \sin 4 \theta b_{111}  \bigg\} (-\beta \ket{0} + \alpha \ket{1}) \Bigg].          
\end{eqnarray}
It is interesting to note that $b_{100}$, $b_{111}$, $b_{113}$, $b_{131}$, $b_{311}$ and $b_{333}$ have both $\beta \ket{0} + \alpha \ket{1}$ and $-\beta \ket{0} + \alpha \ket{1}$ in Bob's last qubit.     
Of course, it reduces to Eq. (\ref{qt2-1}) when $\theta = \pi / 2$.        

\subsection{Bell Measurement of $\{2,3\}$ or $\{1,4\}$ qubits}
In this subsection we assume that Daniel takes  $\{2,3\}$ or $\{1,4\}$ qubits for the Bell measurement.      
Examining Eq. (\ref{qt3-1}) carefully, one can show that the probabilities for outcomes and Bob's $6^{th}$-qubit state are independent of Daniel's choice for measurement.
The probability for each outcome are 
\begin{eqnarray}
\label{prob3-1}
&&P_0 \equiv P(0,0) = \frac{1}{64} \left(36 + 23 \cos 2 \theta + 4 \cos 4 \theta + \cos 6 \theta \right)      \\    \nonumber
&&P_1 \equiv P(0,1) = \frac{\sin^2 \theta}{32} \left(16 + 11 \cos 2 \theta + 4 \cos 4 \theta + \cos 6 \theta \right)     \\   \nonumber
&&P_2 \equiv P(1,0) = \frac{\sin^2 \theta}{16} \left(5 + 2 \cos 2 \theta + \cos 4 \theta \right)                  \\    \nonumber
&&P_3 \equiv P(1,1) = \frac{\sin^4 \theta}{8} \left(7 + 6 \cos 2 \theta + \cos 4 \theta \right).
\end{eqnarray}
It is easy to show $\sum_{i=0}^3 P_i = 1$. After measurement, the Bob's $6^{th}$-qubit state should be derived by taking a partial trace over remaining qubits. Therefore, 
Bob's state can be generally mixed state. In order to examine how well the quantum teleportation is accomplished, we will compute the fidelity ${\cal F} (\rho, \sigma) = \mbox{Tr} \sqrt{\rho^{1/2} \sigma \rho^{1/2}}$
between Alice's secret state and Bob's last-qubit state. If ${\cal F} = 1$, this means a perfect teleportation. 

\begin{center}
\begin{tabular}{c|c|c}  \hline  \hline
 measurement outcome & definition & Bloch vector ${\bf s}$ of Bob's $6^{th}$-qubit state    \\  \hline
                 &                         &     $s_1 = \frac{a + a_-}{P_0}  (\alpha \beta^* + \alpha^* \beta)$  \\     
 $(0,0)$      & $\sigma_{0,B}$ & $s_2 = \frac{i (a_- - a)}{P_0} (\alpha \beta^* - \alpha^* \beta) $        \\
                &                          & $s_3 = \frac{2 a_+ - P_0}{P_0} (|\alpha|^2 - |\beta|^2) $                   \\      \hline
                &                          & $s_1 = - \frac{b_1 + b_2}{P_1}  (\alpha \beta^* + \alpha^* \beta)$   \\
$(0,1)$      & $\sigma_{1,B}$    & $s_2 = - \frac{i (b_1 - b_2)}{P_1} (\alpha \beta^* - \alpha^* \beta) $        \\
               &                           & $s_3 = \frac{2 b_1 - P_1}{P_1}  (|\alpha|^2 - |\beta|^2) $                \\      \hline   
               &                           & $s_1 = -  (\alpha \beta^* + \alpha^* \beta)$                         \\
$(1,0)$     &  $\sigma_{2,B}$  &  $s_2 = - \frac{i (2 c - P_2)}{P_2}  (\alpha \beta^* - \alpha^* \beta) $        \\
              &                            &  $s_3 = \frac{2 c - P_2}{P_2}   (|\alpha|^2 - |\beta|^2) $                   \\      \hline    
              &                            &  $s_1 = - \frac{d_1 + d_2}{P_3}   (\alpha \beta^* + \alpha^* \beta)$  \\
$(1,1)$    & $\sigma_{3,B}$    &  $s_2 = - \frac{i (d_1 - d_2)}{P_3}  (\alpha \beta^* - \alpha^* \beta) $        \\
             &                             &  $s_3 = \frac{2 d_1- P_3}{P_3}  (|\alpha|^2 - |\beta|^2) $                   \\      \hline  \hline

 \end{tabular}
\\
\end{center}

\begin{center}
{\large{Table I}}: Bob's $6^{th}$-qubit state for each measurement outcome. The quantities $a$, $a_{\pm}$, $b_1$, $b_2$, $c$, $d_1$ and $d_2$ are explicitly given in Eq. (\ref{bob3-1}).
\end{center}

\vspace{0.5cm}

%%%%%%%%%%%%%%%%%%%%%%%%%%%%%%%%%%%%%%%%%%%%%%%%%%%%%%%%%
\begin{figure}[ht!]
\begin{center}
\includegraphics[height=5.0cm]{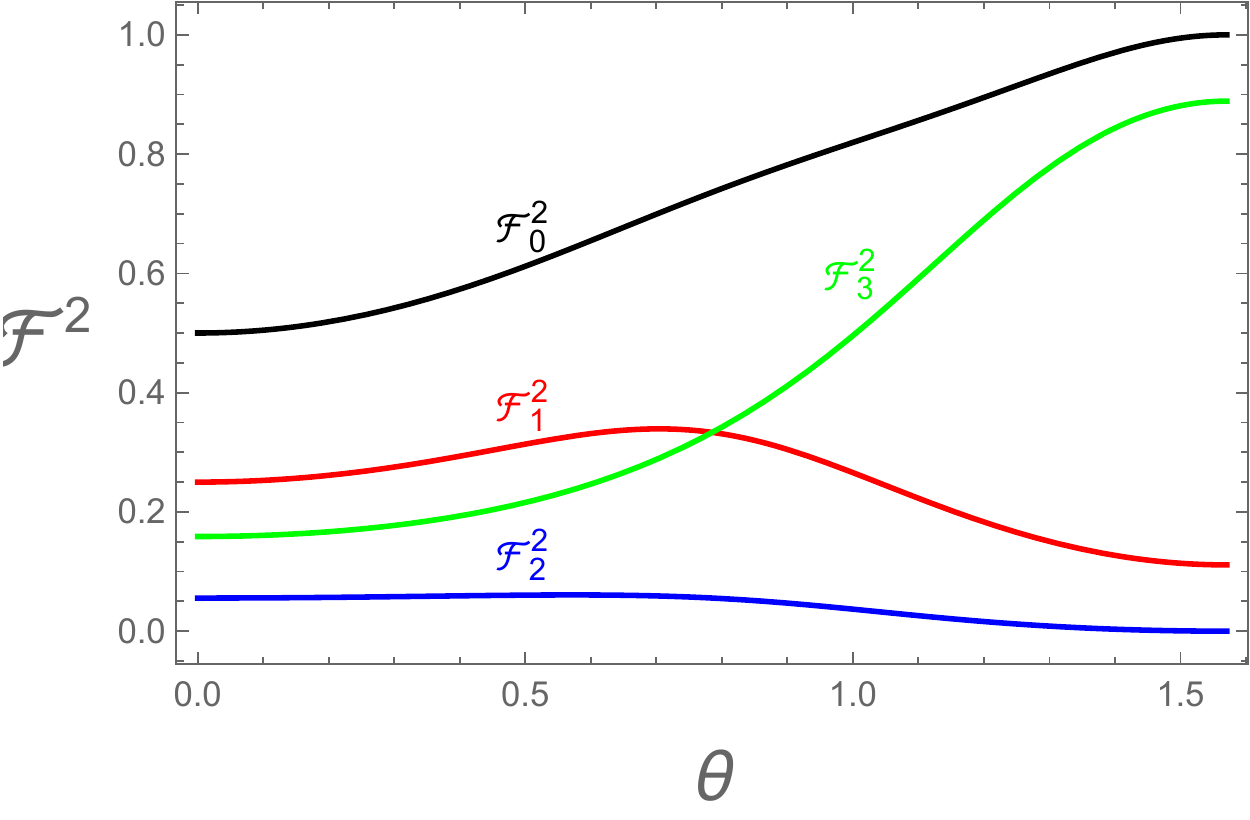} 
\includegraphics[height=5.0cm]{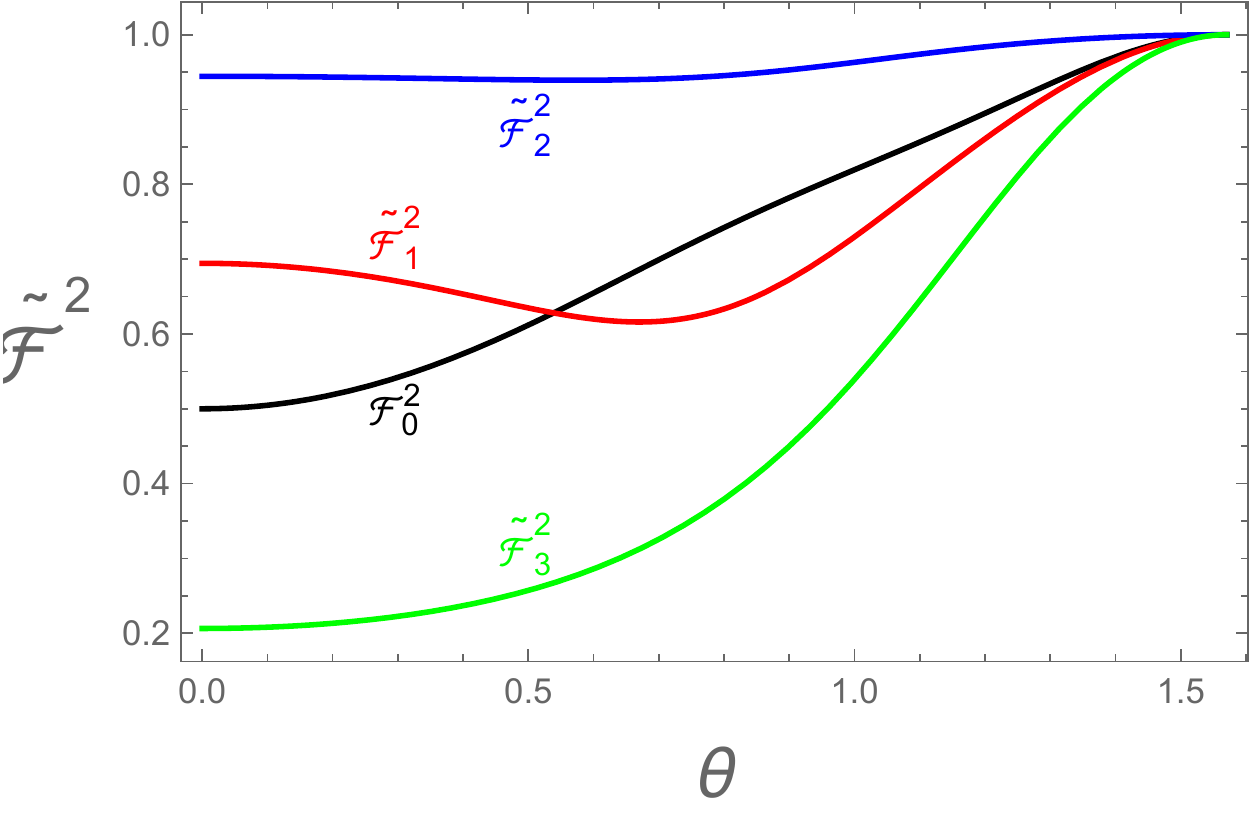}

\caption[fig5]{(Color online) (a) The $\theta$-dependence of ${\cal F}_j^2$  presented in Eq. (\ref{fidel3-2}) when $\alpha = 1 / \sqrt{3}$ and $\phi = 0$.
(b)  The $\theta$-dependence of $\widetilde{{\cal F}}_j^2$  presented in Eq. (\ref{fidel3-3}) when $\alpha = 1 / \sqrt{3}$ and $\phi = 0$.
 }
\end{center}
\end{figure}
%%%%%%%%%%%%%%%%%%%%%%%%%%%%%%%%%%%%%%%%%%%%%%%%%%%%%%%%%%%

In Table I  Bob's $6^{th}$-qubit state is summarized for each measurement outcome, where 
 \begin{eqnarray}
 \label{bob3-1}
&& \hspace{2.0cm}  a = \frac{1}{2} \sin^2 \theta \cos^4 \theta     \hspace{2.0cm}       a_{\pm} = \frac{1}{4} (1 \pm \cos^4 \theta)                            \\    \nonumber
&&  \hspace{2.0cm} b_1 = \frac{\sin^2 \theta}{16} (3 + \cos 4 \theta)   \hspace{2.0cm}  b_2 = \frac{\sin^2 4 \theta}{64}                                                \\    \nonumber
&& c = \frac{\sin^2 2 \theta}{16}    \hspace{1.0cm}  d_1 = \frac{1}{2} \sin^4 \theta \cos^2 \theta   \hspace{1.0cm} d_2 = \frac{\sin^4 \theta}{8} (1 + 4 \cos 2 \theta + \cos 4 \theta).
\end{eqnarray}    
Then, it is straightforward to compute the fidelities ${\cal F}_j^2 = {\cal F}^2 (\rho_A, \sigma_{j,B})$, where $\rho_A = \ket{\psi}_A \bra{\psi}$, whose explicit expressions are in the form:  
\begin{eqnarray}
\label{fidel3-1}
&&{\cal F}_0^2 = \frac{1}{P_0} \left[ a_+ (|\alpha|^4 + |\beta|^4) + 2 (P_0 + a_- - a_+) |\alpha|^2 |\beta|^2 + a \left\{ (\alpha \beta^*)^2 + (\alpha^* \beta)^2 \right\} \right]     \\   \nonumber
&& {\cal F}_1^2 = \frac{1}{P_1} \left[ b_1 (|\alpha|^4 + |\beta|^4) + 2 (P_1 - 2 b_1) |\alpha|^2 |\beta|^2 - b_2 \left\{ (\alpha \beta^*)^2 + (\alpha^* \beta)^2 \right\} \right]     \\   \nonumber
&&{\cal F}_2^2 = \frac{1}{P_2} \left[ c (|\alpha|^4 + |\beta|^4) + 2 (P_2 - 2 c) |\alpha|^2 |\beta|^2 - (P_2 - c) \left\{ (\alpha \beta^*)^2 + (\alpha^* \beta)^2 \right\} \right]     \\   \nonumber
&&{\cal F}_3^2 = \frac{1}{P_3} \left[ d_1 (|\alpha|^4 + |\beta|^4) + 2 (P_3 - 2 d_1) |\alpha|^2 |\beta|^2 - d_2 \left\{ (\alpha \beta^*)^2 + (\alpha^* \beta)^2 \right\} \right].
\end{eqnarray} 
If $\alpha$ is real and $\beta = \sqrt{1 - \alpha^2} e^{i \phi}$, Eq. (\ref{fidel3-1}) becomes  
\begin{eqnarray}
\label{fidel3-2}
&&{\cal F}_0^2 = \frac{1}{P_0} \left[ a_+ \left\{\alpha^4 + (1 - \alpha^2)^2 \right\} + 2  \alpha^2 (1 - \alpha^2) (P_0 + a_- - a_+ + a \cos 2 \phi)  \right]                \\   \nonumber
&&{\cal F}_1^2 = \frac{1}{P_1} \left[ b_1 \left\{\alpha^4 + (1 - \alpha^2)^2 \right\} + 2  \alpha^2 (1 - \alpha^2) (P_1 - 2 b_1 - b_2 \cos 2 \phi)  \right]                \\   \nonumber
&&{\cal F}_2^2 = \frac{1}{P_2} \left[ c \left\{\alpha^4 + (1 - \alpha^2)^2 \right\} + 2  \alpha^2 (1 - \alpha^2) \{P_2 - 2 c - (P_2 - c) \cos 2 \phi \}  \right]                \\   \nonumber
&&{\cal F}_3^2 = \frac{1}{P_3} \left[ d_1 \left\{\alpha^4 + (1 - \alpha^2)^2 \right\} + 2  \alpha^2 (1 - \alpha^2) (P_3 - 2 d_1 - d_2\cos 2 \phi)  \right].
\end{eqnarray}
The $\theta$-dependence of ${\cal F}_j^2$ is plotted in Fig. 5a when $\alpha = 1 / \sqrt{3}$ and $\phi = 0$. This figure shows that ${\cal F}_j^2$ do not reach to $1$ at $\theta = \pi / 2$ except $j = 0$.
In fact, this can be expected from Eq. (\ref{qt2-1}). 

%%%%%%%%%%%%%%%%%%%%%%%%%%%%%%%%%%%%%%%%%%%%%%%%%%%%%%%%%
\begin{figure}[ht!]
\begin{center}
\includegraphics[height=5.0cm]{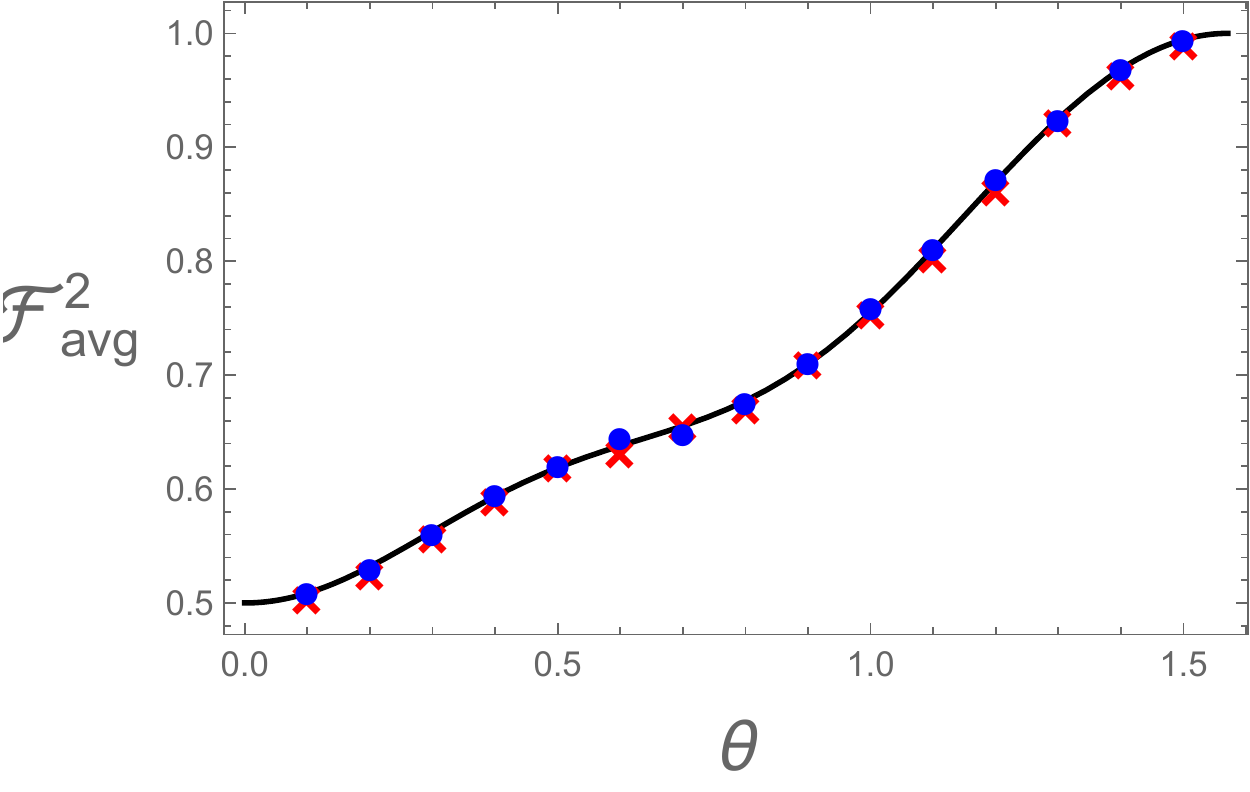} 
\includegraphics[height=5.0cm]{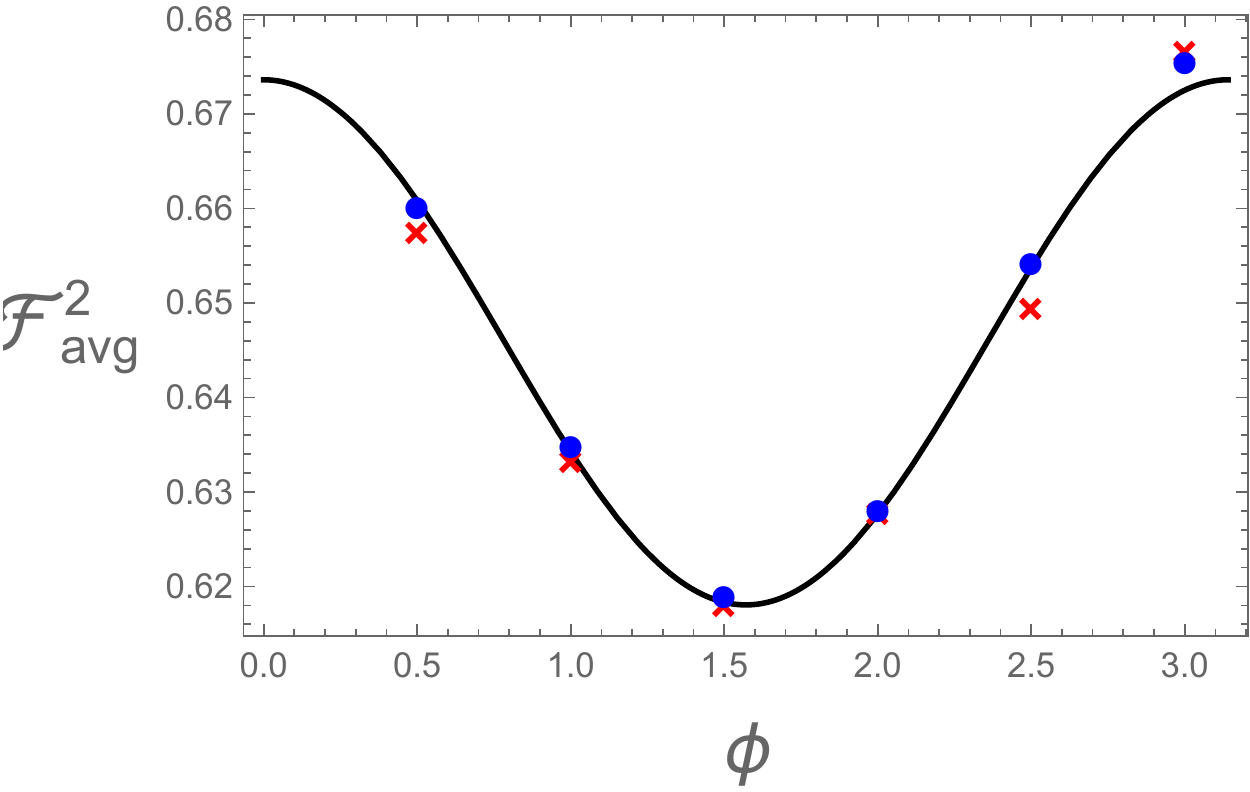}

\caption[fig6]{(Color online) (a)  The $\theta$-dependence of ${\cal F}_{avg}^2$  presented in Eq. (\ref{fidel3-4}) when $\alpha = 1 / \sqrt{3}$ and $\phi = 0$.
(b) The $\phi-$dependence of ${\cal F}_{avg}^2$  when $\alpha = 1 / \sqrt{3}$ and $\theta = \pi / 4$. These figures exhibit monotonic behavior in $\theta$ direction and oscillator behavior in $\phi$ direction.
In both figures the red crossing and blue dot are numerical results computed by qiskit and ibm$\_$oslo, respectively. }
\end{center}
\end{figure}
%%%%%%%%%%%%%%%%%%%%%%%%%%%%%%%%%%%%%%%%%%%%%%%%%%%%%%%%%%%

In order to increase the fidelities at $\theta = \pi / 2$ we define
\begin{equation}
\label{newbob-1}
\widetilde{\sigma}_{1,B} = Z \sigma_{1,B} Z      \hspace{1.0cm}
\widetilde{\sigma}_{2,B} = Z X \sigma_{2,B} X Z    \hspace{1.0cm}
\widetilde{\sigma}_{3,B} = X \sigma_{3,B} X.
\end{equation}
Then, $\widetilde{{\cal F}}_j^2 \equiv {\cal F}^2 (\rho_A, \widetilde{\sigma}_{j,B})$ becomes
\begin{eqnarray}
\label{fidel3-3}
&& \widetilde{{\cal F}}_1^2 = \frac{1}{P_1} \left[ b_1 (|\alpha|^4 + |\beta|^4) + 2 P_1 |\alpha|^2 |\beta|^2 + b_2 \left\{ (\alpha \beta^*)^2 + (\alpha^* \beta)^2 \right\} \right]    \\    \nonumber
&& \hspace{0.7cm} = \frac{1}{P_1} \left[ b_1 \left\{\alpha^4 + (1 - \alpha^2)^2 \right\}  +  2  \alpha^2 (1 - \alpha^2) (P_1 + b_2 \cos 2 \phi) \right]                                                          \\   \nonumber
&& \widetilde{{\cal F}}_2^2 = \frac{1}{P_2} \left[ (P_2 - c) (|\alpha|^4 + |\beta|^4) + 2 P_2 |\alpha|^2 |\beta|^2 + c \left\{ (\alpha \beta^*)^2 + (\alpha^* \beta)^2 \right\} \right]    \\    \nonumber
&& \hspace{0.7cm} = \frac{1}{P_2} \left[ (P_2 - c) \left\{\alpha^4 + (1 - \alpha^2)^2 \right\}  +  2  \alpha^2 (1 - \alpha^2) (P_2 + c \cos 2 \phi) \right]                                                          \\   \nonumber
&& \widetilde{{\cal F}}_3^2 = \frac{1}{P_3} \left[ (P_3 - d_1) (|\alpha|^4 + |\beta|^4) + 2 (d_1 - d_2) |\alpha|^2 |\beta|^2 - d_1 \left\{ (\alpha \beta^*)^2 + (\alpha^* \beta)^2 \right\} \right]  \\   \nonumber
&& \hspace{0.7cm} = \frac{1}{P_3} \left[ (P_3 - d_1) \left\{\alpha^4 + (1 - \alpha^2)^2 \right\}  +  2  \alpha^2 (1 - \alpha^2) (d_1 - d_2 - d_1 \cos 2 \phi) \right].                                                     
\end{eqnarray}
The $\theta$-dependence of  $\widetilde{{\cal F}}_j^2$ is plotted in Fig. 5b when $\alpha = 1 / \sqrt{3}$ and $\phi = 0$. As expected, this figure shows that all $\widetilde{{\cal F}}_j^2$ approach to $1$
at $\theta = \pi / 2$, which indicates the perfect teleportation in the maximal scrambling unitary (\ref{scram2-1}). In Fig. 6a we plot the $\theta$-dependence of the average fidelity defined 
\begin{equation}
\label{fidel3-4}
{\cal F}_{avg}^2 = P_0 {\cal F}_0^2 + P_1 \widetilde{{\cal F}}_1^2 + P_2 \widetilde{{\cal F}}_2^2 + P_3 \widetilde{{\cal F}}_3^2
\end{equation}
when $\alpha = 1 / \sqrt{3}$ and $\phi = 0$. It approaches to $0.5$ and $1$ when $\theta = 0$ (no scrambling) and $\theta = \pi / 2$ (maximal scrambling). 
The monotonic increasing behavior of ${\cal F}_{avg}^2$ supports the conjecture ``the quantity of scrambling is proportional to the fidelity of quantum teleportation''. 
 In Fig. 6b we plot the $\phi$-dependence of ${\cal F}_{avg}^2$ when $\alpha = 1 / \sqrt{3}$ and $\theta = \pi / 4$. As expected, this figure 
exhibits oscillatory behavior. In Fig 6 the red crossing and blue dot are numerical results computed by qiskit and ibm$\_$oslo.
This will be discussed in next section. 

%\subsection{Bell Measurement of $(1,4)$ qubits}
%Now, let assume that Bob performs a Bell measurement on $(1,4)$ qubits
%Eq. (\ref{qt3-1}) implies that the Bob's last state $\sigma_{j,B}$ is exactly the same with the case of previous subsection. 
%Thus, all fidelities are exactly the same with the case of ``Bell Measurement of $(2,3$ qubits''.
%Eq. (\ref{qt2-1}) implies that the perfect teleportation is possible at $\theta = \pi / 2$ even though Bob performs a Bell measurement on $(0,5)$ qubits. 
%Of course, this is mathematically possible. Since, however, Bob cannot accesses to Alice, this possibility is meaningless physically. 
%In spite of the fact we address this problem in appendix B for completeness. 

\subsection{Bell Measurement of $\{0,5\}$qubits}
%%%%%%%%%%%%%%%%%%%%%%%%%%%%%%%%%%%%%%%%%%%%%%%%%%%%%%%%%
\begin{figure}[ht!]
\begin{center}
\includegraphics[height=5.0cm]{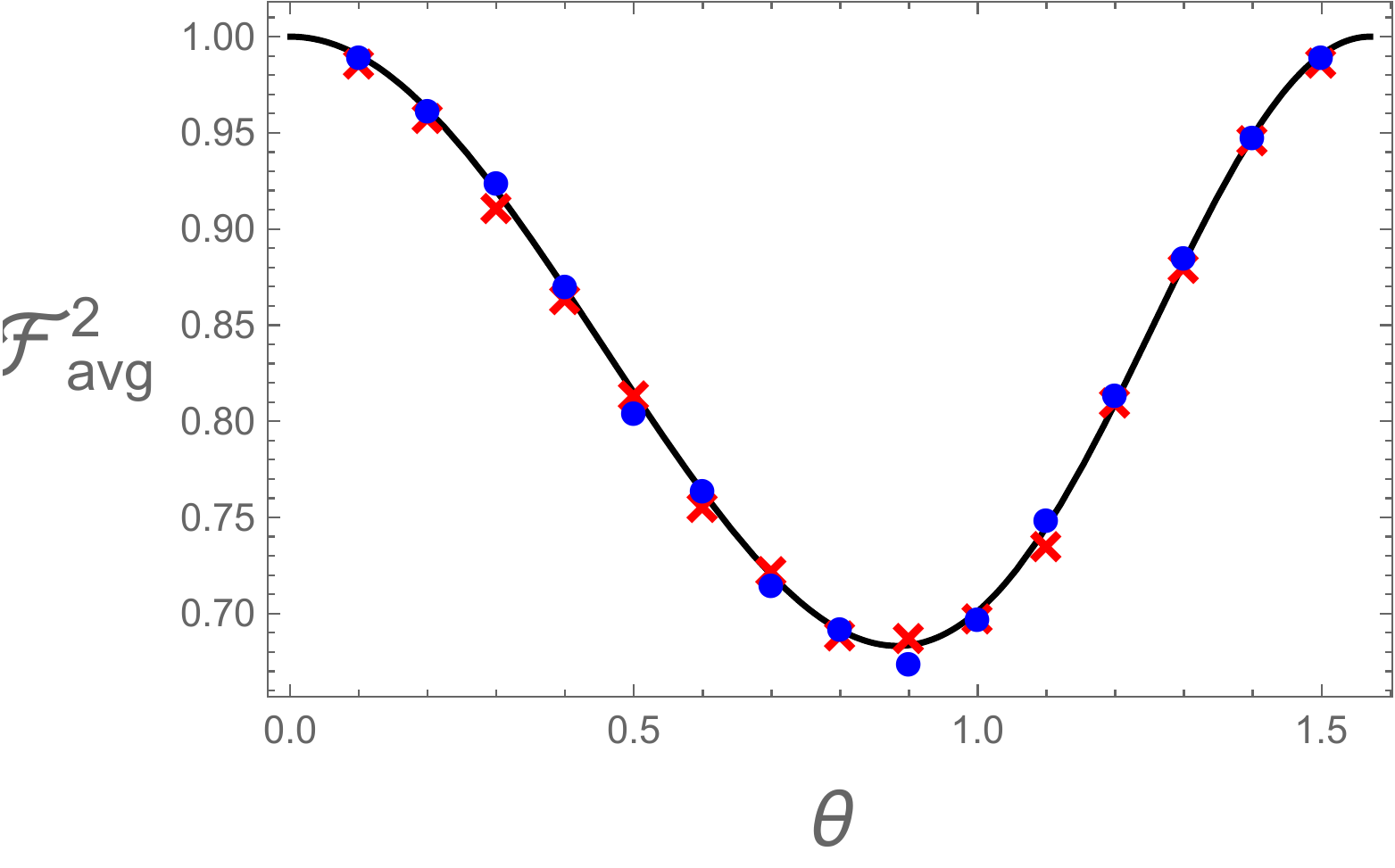} 
\includegraphics[height=5.0cm]{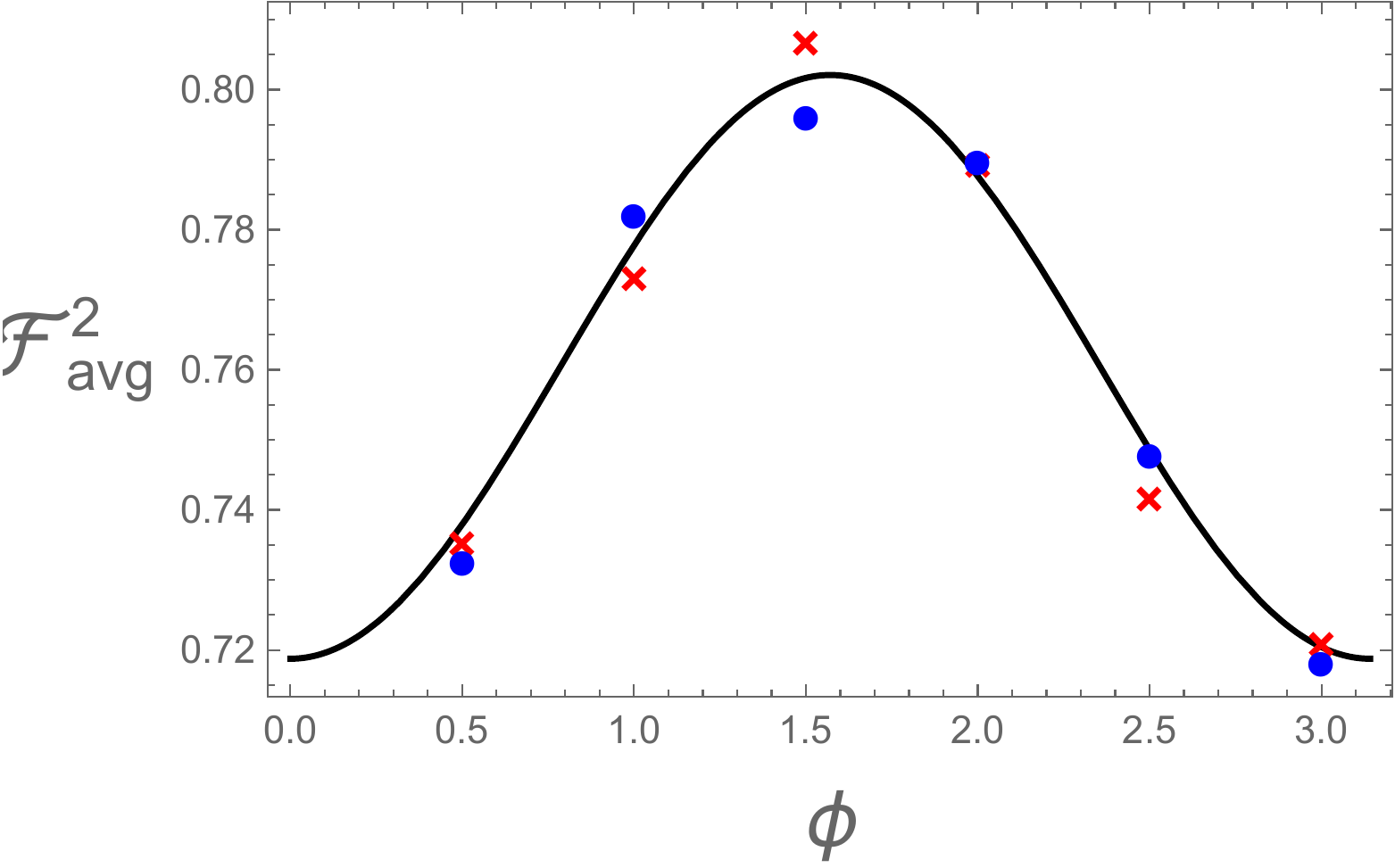}

\caption[fig7]{(Color online) (a)  The $\theta$-dependence of ${\cal F}_{avg}^2$  presented in Eq. (\ref{app-fidel3-3}) when $\alpha = 1 / \sqrt{3}$ and $\phi = 0$.
(b) The $\phi-$dependence of ${\cal F}_{avg}^2$  when $\alpha = 1 / \sqrt{3}$ and $\theta = \pi / 3$. In both figures the red crossing and blue dot are numerical results  computed by qiskit and ibm$\_$oslo, respectively.
 }
\end{center}
\end{figure}
%%%%%%%%%%%%%%%%%%%%%%%%%%%%%%%%%%%%%%%%%%%%%%%%%%%%%%%%%%%

In this subsection we assume that Daniel takes  $\{0,5\}$ qubits as a Bell measurement. From Eq. (\ref{qt3-1}) it is straightforward to show that 
the probability $Q_j$ for each outcome is  
\begin{eqnarray}
\label{app-prob3-1}
&&Q_0 \equiv P(0,0) = \frac{1}{4} \left(1 + 2 \sin^4 \theta \cos ^2 \theta  \right)      \\    \nonumber
&&Q_1 \equiv P(0,1) = \frac{1}{128} \left(33 -2 \cos 2 \theta + 2 \cos 6\theta - \cos 8 \theta \right)     \\   \nonumber
&&Q_2 \equiv P(1,0) = \frac{1}{4} \left(\cos^4 \theta + \sin^8 \theta + 2\sin^2 \theta \cos^2 \theta -\sin^4 \theta \cos^4 \theta \right)                  \\    \nonumber
%&&P_2' \equiv P'(1,0) =\frac{1}{64} \left(14 + \cos 2 \theta + 2 \cos 4 \theta - \cos 6 \theta \right)                  \\    \nonumber
&&Q_3 \equiv P(1,1) =  \frac{1}{128} \left(31 +2 \cos 2 \theta - 2 \cos 6\theta + \cos 8 \theta \right).
\end{eqnarray}

\begin{center}
\begin{tabular}{c|c|c}  \hline  \hline
 measurement outcome & definition & Bloch vector ${\bf s}$ of Bob's final state    \\  \hline
                 &                         &     ${s}_1 = \frac{1-2Q_0}{2Q_0}  (\alpha \beta^* + \alpha^* \beta)$  \\     
 $(0,0)$      & $\rho_{0,B}$ & ${s}_2 = i (\alpha \beta^* - \alpha^* \beta) $        \\
                &                          & ${s}_3 = \frac{1-2Q_0}{2Q_0} (|\alpha|^2 - |\beta|^2) $                   \\      \hline
                &                          & $s_1 =  \frac{1}{Q_1}  [\left (y_1 -x_1)(\alpha \beta^* + \alpha^* \beta) - i z_1 (\alpha \beta^* - \alpha^* \beta)   \right]$   \\
$(0,1)$      & $\rho_{1,B}$    & $s_2 = -\frac{i}{Q_1}  [\left (y_1 + x_1)(\alpha \beta^* - \alpha^* \beta) - i z_1 (\alpha \beta^* + \alpha^* \beta)   \right]$   \\
               &                           & $s_3 = \frac{ 2 x_1 -Q_1}{Q_1}  (|\alpha|^2 - |\beta|^2) $                \\      \hline   
               &                           & $s_1 =   -(\alpha \beta^* + \alpha^* \beta)$                         \\
$(1,0)$     &  $\rho_{2,B}$  &  $s_2 =  -\frac{i ( Q_2 -2 x_2 )}{Q_2}  (\alpha \beta^* - \alpha^* \beta) $        \\
              &                            &  $s_3 = -\frac{2x_2 -Q_2}{Q_2}   (|\alpha|^2 - |\beta|^2) $                   \\      \hline    
              &                            &  $s_1 =\frac{1}{Q_3}  [\left (y_3 -x_3)(\alpha \beta^* + \alpha^* \beta) - i z_3 (\alpha \beta^* - \alpha^* \beta)   \right]$   \\
$(1,1)$    & $\rho_{3,B}$    &  $s_2 = -\frac{i}{Q_3}  [\left (y_3 +x_3)(\alpha \beta^* - \alpha^* \beta) - i z_3 (\alpha \beta^* + \alpha^* \beta)   \right]$   \\
             &                             &  $s_3 = \frac{2 x_3 - Q_3}{Q_3}  (|\alpha|^2 - |\beta|^2) $                   \\      \hline  \hline
 
 \end{tabular}
\\
\end{center}
\begin{center}
{\large{Table II}}: Bob's $6^{th}$-qubit state. The quantities $x_1$,$x_2$, $x_3$,, $y_1$, $y_3$, $z_1$ and $z_3$ are explicitly given in Eq. (\ref{app-bob3-1}).
\end{center}
\vspace{0.5cm}

After taking a partial trace over remaining qubits, Bob's $6^{th}$-qubit state, $\rho_{j, B}$, for each measurement outcome is summarized in Table II, where 
\begin{eqnarray}
\label{app-bob3-1}
&& \hspace{1cm}  x_1 = \frac{1}{2} \sin^4 \theta \cos^2 \theta     \hspace{2.3cm}  x_2 =x_3 = \frac{1}{2} \sin^2 \theta \cos^4 \theta         \\    \nonumber                       
&&  \hspace{1cm}   y_1 = \frac{1}{64} (3+ \cos 4 \theta)^2      \hspace{2.0cm}                                                  y_3 = -\frac{1}{128}(11+20 \cos 4 \theta + \cos 8 \theta)  \\    \nonumber
&& \hspace{1cm}  z_1 = \frac{\cos^3 \theta}{4} (\sin \theta - \sin 3 \theta)  \hspace{1cm}  z_3 = \frac{1}{8} \sin 4 \theta  \cos ^2 \theta .
\end{eqnarray} 

In order to increase the fidelities at $\theta = \pi / 2$ we define
\begin{equation}
\label{app-newbob-1}
\widetilde{\rho}_{1,B} = X \rho_{1,B} X      \hspace{1.0cm}
\widetilde{\rho}_{2,B} = Z \rho_{2,B} Z      \hspace{1.0cm}
\widetilde{\rho}_{3,B} = Z X \rho_{3,B} X Z.
\end{equation}
It is interesting to note that $\rho_{0, B}$ and $\widetilde{\rho}_{j, B} \hspace{0.2cm} (j=1,2,3)$ reduce to $\rho_A = \ket{\psi}_A \bra{\psi}$ at $\theta = 0$ and $\pi/2$. 
Therefore, the fidelities ${\cal F}_0^2 \equiv {\cal F}^2 (\rho_A, \rho_{0,B})$ and  $\widetilde{{\cal F}}_j^2 \equiv {\cal F}^2 (\rho_A, \widetilde{\rho}_{j,B})$ should be one at both $\theta = 0$ and $\pi/2$.
The explicit expressions of those fidelities are 
 \begin{eqnarray}
\label{app-fidel3-3}
&&{\cal F}_0^2 = \frac{1}{4 Q_0} \left[|\alpha|^4 + |\beta|^4 + 8 Q_0 |\alpha|^2 |\beta|^2 + (1-4Q_0 ) \left\{ (\alpha \beta^*)^2 + (\alpha^* \beta)^2 \right\} \right]     \\   \nonumber
&& \hspace{0.7cm} = \frac{1}{4 Q_0} \left[ \alpha^4 + (1 - \alpha^2)^2 +2  \alpha^2 (1 - \alpha^2)  \{ 4Q_0 + (1-4Q_0 )\cos 2 \phi \} \right]                \\   \nonumber
&& \widetilde{{\cal F}}_1^2 = \frac{1}{Q_1} \left[ (Q_1 - x_1 ) (|\alpha|^4 + |\beta|^4) + 2( x_1 +y_1 )|\alpha|^2 |\beta|^2 - x_1 \left\{ (\alpha \beta^*)^2 + (\alpha^* \beta)^2 \right\} \right]    \\    \nonumber
&& \hspace{0.7cm} = \frac{1}{Q_1} \left[ (Q_1 - x_1 )  \left\{\alpha^4 + (1 - \alpha^2)^2 \right\}  +  2  \alpha^2 (1 - \alpha^2) (-x_1 \cos 2 \phi +x_1 +y_1) \right]                                                          \\   \nonumber
&& \widetilde{{\cal F}}_2^2 = \frac{1}{Q_2} \left[ (Q_2 - x_2) (|\alpha|^4 + |\beta|^4) + 2 Q_2 |\alpha|^2 |\beta|^2 + x_2 \left\{ (\alpha \beta^*)^2 + (\alpha^* \beta)^2 \right\} \right]    \\    \nonumber
&& \hspace{0.7cm} =\frac{1}{Q_2} \left[ (Q_2 - x_2) \left\{\alpha^4 + (1 - \alpha^2)^2 \right\}  +  2  \alpha^2 (1 - \alpha^2)(x_2  \cos 2 \phi + P_2 ) \right]                                                          \\   \nonumber
&& \widetilde{{\cal F}}_3^2 =  \frac{1}{Q_3} \left[ (Q_3 - x_3 ) (|\alpha|^4 + |\beta|^4) + 2( x_3 -y_3 )|\alpha|^2 |\beta|^2 + x_3 \left\{ (\alpha \beta^*)^2 + (\alpha^* \beta)^2 \right\} \right]    \\    \nonumber
&& \hspace{0.7cm} = \frac{1}{Q_3} \left[ (Q_3 - x_3 )  \left\{\alpha^4 + (1 - \alpha^2)^2 \right\}  +  2  \alpha^2 (1 - \alpha^2) (x_3 \cos 2 \phi +x_3 -y_3) \right],                                                    
\end{eqnarray}
where $\beta = \sqrt{1 - \alpha^2} e^{i \phi}$ for real $\alpha$ is used in the second expression of each fidelity.

In Fig. 7a we plot the $\theta$-dependence of the average fidelity defined by Eq. (\ref{fidel3-4}) when $\alpha = 1 / \sqrt{3}$ and $\phi = 0$.
As expected it approaches to $1$ at both $\theta = 0$ and $\theta = \pi / 2$, which does not support the conjecture ``the quantity of scrambling is proportional to the fidelity of quantum teleportation''.
In Fig. 7b the $\phi$-dependence of average fidelity is plotted when  $\alpha = 1 / \sqrt{3}$ and $\theta = \pi / 3$. 
In Fig 7 the red crossing and blue dot are numerical results computed by qiskit and ibm$\_$oslo.

\section{Numerical Simulation}

%%%%%%%%%%%%%%%%%%%%%%%%%%%%%%%%%%%%%%%%%%%%%%%%%%%%%%%%%
\begin{figure}[ht!]
\begin{center}
\includegraphics[height=6.0cm]{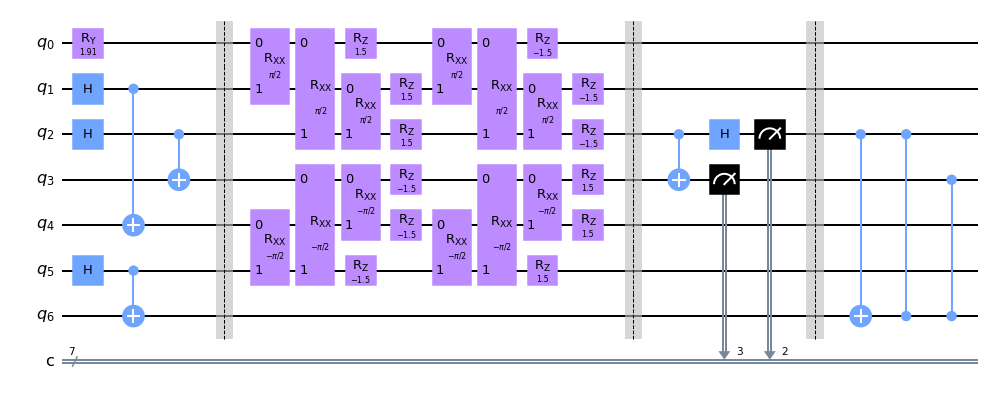} 
\includegraphics[height=6.0cm]{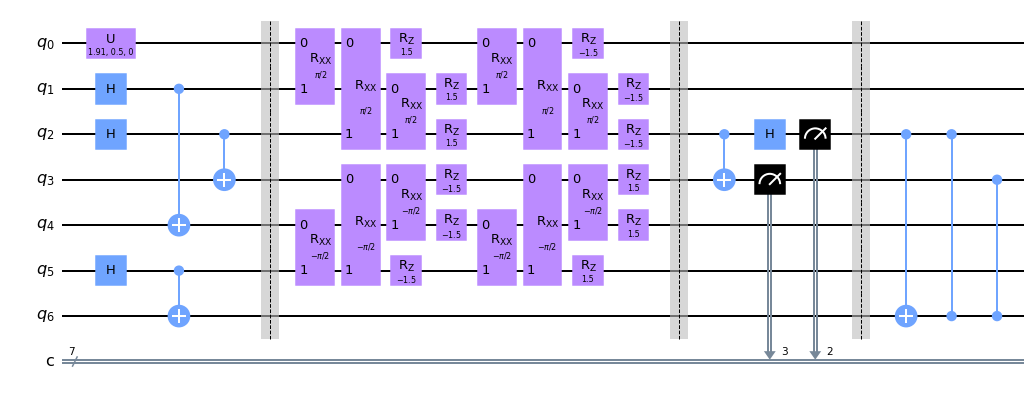}

\caption[fig8]{(Color online) (a) Full quantum circuit for the teleportation when $\ket{\psi}_A = \frac{1}{\sqrt{3}} (\ket{0} + \sqrt{2} \ket{1})$ when $\theta = 1.5$ and $\phi = 0$.
(b) Full quantum circuit for the teleportation when $\ket{\psi}_A = \frac{1}{\sqrt{3}} (\ket{0} + \sqrt{2} e^{i \phi} \ket{1})$ when $\theta = 1.5$ and $\phi = 0.5$.
In both figures it is assumed that Daniel chooses $\{2,3\}$ qubits for a Bell measurement. }
\end{center}
\end{figure}
%%%%%%%%%%%%%%%%%%%%%%%%%%%%%%%%%%%%%%%%%%%%%%%%%%%%%%%%%%%

In order to examine the noise effect we compute the fidelities numerically in this section by making use of the qiskit and $7$-qubit real quantum computer ibm$\_$oslo, and compare them with the theoretical results. 
First, we assume that Alice's secret state is $\ket{\psi}_A = \sqrt{\frac{1}{3}} \ket{0} + \sqrt{\frac{2}{3}} \ket{1}$. In order to compute the fidelities numerically we prepare a quantum 
circuit of Fig. 8a.  In this figure $\theta=1.5$ is chosen and we assume that Daniel chooses $\{2,3\}$ qubits for  Bell measurement. 
In the circuit the gates of purple color represent $U$ and $U^*$ presented in Eq. (\ref{scram3-1}). The numerical experiment is repeated $10^3$ times and we compute their average value. 

Next, we take $\ket{\psi}_A = \sqrt{\frac{1}{3}} \ket{0} + \sqrt{\frac{2}{3}} e^{i \phi} \ket{1}$. In this case we should prepare a quantum circuit of Fig. 8b. In this figure we choose $\theta=1.5$ and $\phi=0.5$, and Bell measurement of $\{2,3\}$ qubits.
The numerical results are summarized in appendix B and C as Table IV, V, VI, and VII.

\begin{center}
\begin{tabular}{|c|c|c|c|c|}       \hline  \hline
                          & \multicolumn{2} {c|} { qiskit}  &  \multicolumn{2} {c|} {ibm$\_$oslo}          \\  \hline
$\{2,3\}$ or $\{1,4\}$ qubits & Table IV   &  Table V  &  Table IV  &  Table V                                                            \\   \cline{2-5}
Bell measurement   &   $0.225$    &   $0.190$    &  $0.219$   &  $0.114$                                                      \\   \hline   \hline
$\{0,5\}$ qubits     &  Table VI     &    Table VII    &   Table VI   &   Table VII                                                      \\   \cline{2-5}
Bell Measurement  &  $0.279$   &  $0.360$    &   $0.329$   &   $0.401$                                                       \\   \hline  \hline
 \end{tabular}
\\
\end{center}

\begin{center}
{\large{Table III}}: Error of qiskit and  ibm$\_$oslo in Table IV, V, VI, and VII. 
\end{center}
\vspace{0.5cm}

If we define the error as $ \bigg(\frac{1}{N} \sum (\mbox{theoretical value} - \mbox{experimental value}) \bigg) \times 100$ where $N$ is a number of data, the errors derived from Table IV, V, VI, and VII are summarized in Table III.
From the Table the noise effect is negligible in the computation of the average fidelities although the  ibm$\_$oslo is little bit robust than qiskit against noise at Table IV and V, and vice versa at Table VI and VII . 
Thus, we can adopt the numerical approach when we need to discuss a similar issue with huge number of qubits, where analytical calculation of fidelities is highly difficult.

\section{Conclusions}
In this paper we study the role of scrambling unitary in the quantum teleportation scheme. 
In order to explore the issue we introduce $U(\theta)$ in Eq. (\ref{scram3-1}), which parametrizes identity and maximally scrambling unitary (\ref{scram2-1}) 
at $\theta = 0$ and $\theta = \pi / 2$, respectively. Of course, it is a partially scrambling unitary when $0 < \theta < \pi / 2$.

Applying $U(\theta)$ to the $7$-qubit quantum circuit presented in Fig. 1b, we compute the fidelities between Alice's secret state $\alpha \ket{0} + \beta \ket{1}$ and Bob's $6^{th}$-qubit state
after Bell measurement in one of $\{2,3\}$, $\{1,4\}$, or $\{0,5\}$ qubits. 

For the case of Bell measurement of $\{2,3\}$ or $\{1,4\}$ qubits, it is shown that the fidelities are exactly the same. The average fidelity exhibits a monotonic behavior from $0.5$ to $1$ in $\theta$, which supports our conjecture ``the quantity of scrambling is proportional to the fidelity of quantum teleportation''.
For the case of Bell measurement of $\{0,5\}$ qubits, however, perfect teleportation occurs at $\theta = 0$ and $\theta = \pi / 2$. 
Thus, in this case the result does not support the conjecture.
In this reason we conclude that the proportionality of scrambling with fidelity is dependent on the Daniel's choice of qubits for the Bell measurement. 
If $\alpha$ is real and $\beta = \sqrt{1 - \alpha^2} e^{i \phi}$, the average fidelities exhibit an oscillatory behavior in $\phi$.
All fidelities are compared to the numerical results computed by qiskit and ibm$\_$oslo. It is shown that the noise effect is negligible when quantum computer is used to compute the average fidelity.
Therefore, we can adopt the numerical approach when we discuss a similar issue with a quantum circuit of large number of qubits, where analytical calculation of fidelities is highly difficult.

In this paper it is shown that perfect teleportation is possible if $U$ is maximally scrambling unitary as shown in Eq. (\ref{scram2-1}). 
However, there exist so many maximally scrambling unitary. 
We are not sure whether all maximal scrambling always allow a perfect teleportation or not. 
We want to examine this issue in the future. 

So far, we examined the role of scrambling in the teleportation scheme from the aspect of pure QIT. 
However, our analysis has some implication in the information loss problem if we replace Charlie' qubits with black hole's qubits and Daniel's qubits with qubits of Bob's quantum memory.
If we model  the internal dynamics of a black hole by fast scrambling random unitary, Fig. 1b can be interpreted as a quantum teleportation in the black hole. 
Thus, the role of $U$ in Fig. 1b is to mix the quantum information of Alice's and black hole's qubits. 
The Bell measurement corresponds to the measurement of Hawking quanta. 
Then, Eq. (\ref{qt2-1}) implies that the complete decoding of Alice's secret state is possible if $U$ is maximally scrambling unitary. 
In this case we are not sure how asymptotic observer Bob can get a information on black hole's random unitary $U$. Without the information how can 
he apply $U^*$ to his quantum memory and ancillary qubits. To be honest, we have no definite answer on this question.

{\bf Acknowledgement}:
This work was supported by the National Research Foundation of Korea(NRF) grant funded by the Korea government(MSIT) (No. 2021R1A2C1094580).

\newpage 

\begin{appendix}{\centerline{\bf Appendix A: Partial scrambling property of Eq. (\ref{scram3-1}) }}

\setcounter{equation}{0}
\renewcommand{\theequation}{A.\arabic{equation}}

In this appendix we summarize how the maximal scrambling property (\ref{scram2-2}) is changed by the partial scrambling unitary (\ref{scram3-1}). After long calculation,one can show 
\begin{eqnarray}
\label{app-1}
&&U^{\dagger} (X \otimes I \otimes I) U = -\sin^3 \theta \cos \theta (I \otimes I \otimes Y + I \otimes Y \otimes I + Y \otimes Y \otimes Y)                                        \\   \nonumber  
&& + \sin^2 \theta \cos^2 \theta (-X \otimes X \otimes X + Z \otimes X \otimes Z + Z \otimes Z \otimes X) + \sin \theta \cos^3 \theta (Y \otimes I \otimes I)            \\   \nonumber  
&& - \sin^4 \theta (X \otimes Z \otimes Z) + \sin \theta \cos \theta (Y \otimes X \otimes X) + \cos^2 \theta (X \otimes I \otimes I)                                                    \\   \nonumber   
&&U^{\dagger} (I \otimes X \otimes I) U = -\sin^3 \theta \cos \theta (I \otimes I \otimes Y + Y \otimes I \otimes I + Y \otimes Y \otimes Y)                                         \\   \nonumber   
&& + \sin^2 \theta \cos^2 \theta (-X \otimes X \otimes X + X \otimes Z \otimes Z + Z \otimes Z \otimes X) + \sin \theta \cos^3 \theta (I \otimes Y \otimes I)           \\    \nonumber   
&& - \sin^4 \theta (Z \otimes X \otimes Z) + \sin \theta \cos \theta (X \otimes Y \otimes X) + \cos^2 \theta (I \otimes X \otimes I)                                                    \\    \nonumber   
&&U^{\dagger} (I \otimes I \otimes X) U = -\sin^3 \theta \cos \theta (I \otimes Y \otimes I + Y \otimes I \otimes I + Y \otimes Y \otimes Y)                                          \\   \nonumber
&& + \sin^2 \theta \cos^2 \theta (-X \otimes X \otimes X + X \otimes Z \otimes Z + Z \otimes X \otimes Z) + \sin \theta \cos^3 \theta (I \otimes I \otimes Y)             \\   \nonumber
&& - \sin^4 \theta (Z \otimes Z \otimes X) + \sin \theta \cos \theta (X \otimes X \otimes Y) + \cos^2 \theta (I \otimes I \otimes X)                                                     \\   \nonumber
&&U^{\dagger} (Y \otimes I \otimes I) U = -\sin^2 \theta \cos^2 \theta (I \otimes I \otimes Y + I \otimes Y \otimes I + Y \otimes Y \otimes Y)                                          \\   \nonumber
&& + \sin \theta \cos^3 \theta (-X \otimes X \otimes X + Z \otimes X \otimes Z + Z \otimes Z \otimes X) - \sin^3 \theta \cos \theta (X \otimes Z \otimes Z)             \\   \nonumber
&& + \cos^4 \theta (Y \otimes I \otimes I) - \sin \theta \cos \theta (X \otimes I \otimes I) - \sin^2 \theta (Y \otimes X \otimes X)                                                     \\   \nonumber
&&U^{\dagger} (I \otimes Y \otimes I) U = -\sin^2 \theta \cos^2 \theta (I \otimes I \otimes Y + Y \otimes I \otimes I + Y \otimes Y \otimes Y)                                          \\   \nonumber
&& + \sin \theta \cos^3 \theta (-X \otimes X \otimes X + X \otimes Z \otimes Z + Z \otimes Z \otimes X) - \sin^3 \theta \cos \theta (Z \otimes X \otimes Z)             \\   \nonumber
&& + \cos^4 \theta (I \otimes Y \otimes I) - \sin \theta \cos \theta (I \otimes X \otimes I) - \sin^2 \theta (X \otimes Y \otimes X)                                                     \\   \nonumber
&&U^{\dagger} (I \otimes I \otimes Y) U = -\sin^2 \theta \cos^2 \theta (I \otimes Y \otimes I + Y \otimes I \otimes I + Y \otimes Y \otimes Y)                                          \\   \nonumber
&& + \sin \theta \cos^3 \theta (-X \otimes X \otimes X + X \otimes Z \otimes Z + Z \otimes X \otimes Z) - \sin^3 \theta \cos \theta (Z \otimes Z \otimes X)             \\   \nonumber
&& + \cos^4 \theta (I \otimes I \otimes Y) - \sin \theta \cos \theta (I \otimes I \otimes X) - \sin^2 \theta (X \otimes X \otimes Y)                                                     \\   \nonumber
&&U^{\dagger} (Z \otimes I \otimes I) U = \cos^2 \theta (Z \otimes I \otimes I) - \sin^2 \theta (Z \otimes Y \otimes Y)                                                                        \\   \nonumber
&&\hspace{4.0cm}- \sin \theta \cos \theta (Y \otimes X \otimes Z + Y \otimes Z \otimes X)                                                                                                                                             \\   \nonumber
&&U^{\dagger} (I \otimes Z \otimes I) U = \cos^2 \theta (I \otimes Z \otimes I) - \sin^2 \theta (Y \otimes Z \otimes Y)                                                                        \\   \nonumber
&&\hspace{4.0cm}- \sin \theta \cos \theta (X \otimes Y \otimes Z + Z \otimes Y \otimes X)                                                                                                                                             \\   \nonumber
&&U^{\dagger} (I \otimes I \otimes Z) U = \cos^2 \theta (I \otimes I \otimes Z) - \sin^2 \theta (Y \otimes Y \otimes Z)                                                                        \\   \nonumber
&&\hspace{4.0cm}- \sin \theta \cos \theta (X \otimes Z \otimes Y + Z \otimes X \otimes Y).                                                                                                                                            
 \end{eqnarray}
 It is straightforward to show that Eq. (\ref{app-1}) reduces to the maximal scrambling property (\ref{scram2-2}) when $\theta = \pi / 2$.

\end{appendix}

\newpage 

\begin{appendix}{\centerline{\bf Appendix B: Numerical result for Bell Measurement of $\{2,3\}$ or $\{1,4\}$ qubits }}
\setcounter{equation}{0}
\renewcommand{\theequation}{B.\arabic{equation}}

\begin{center}
\begin{tabular}{c|c|c|c}  \hline  \hline
$\theta$ & Fidelity (qiskit Exp.)  & Fidelity (ibm$\_$oslo EXP.) & Fidelity (Theory)   \\  \hline
$0.1$  &  $0.50874$  & $0.50904$  &  $0.50865$    \\
$0.2$  &  $0.52910$  & $0.53068$  &  $0.53186$    \\
$0.3$  &  $0.56235$  & $0.56137$  &   $0.56276$   \\
$0.4$  &  $0.59419$  & $0.59487$ &  $0.59347$    \\
$0.5$  &  $0.62383$  & $0.62094$ &  $0.61871$    \\
$0.6$  &  $0.63689$  & $0.64570$  &  $0.63794$    \\
$0.7$  &  $0.66061$  & $0.64856$ &  $0.65532$    \\
$0.8$  &  $0.67556$  & $0.67637$  &  $0.67737$    \\
$0.9$  &  $0.71438$  & $0.71070$  &  $0.70980$    \\
$1.0$  &  $0.75879$  & $0.75926$  &  $0.75483$    \\
$1.1$  &  $0.80762$  & $0.81087$  &  $0.81011$    \\
$1.2$  &  $0.86660$  & $0.87261$  &  $0.86950$    \\
$1.3$  &  $0.92697$  & $0.92419$  &  $0.92494$    \\
$1.4$  &  $0.96811$  & $0.96934$  &  $0.96861$    \\
$1.5$  &  $0.99451$  & $0.99458$  &  $0.99446$    \\   \hline  \hline
 \end{tabular}
\\
\end{center}

\begin{center}
{\large{Table IV}}: Experimental and Theoretical Fidelities when $\ket{\psi}_A = \sqrt{1/3} \ket{0} + \sqrt{2/3} \ket{1}$.
\end{center}

\vspace{0.5cm}

%Next, we assume that the Alice's secret state is $\ket{\psi}_A = \sqrt{\frac{1}{3}} \ket{0} + \sqrt{\frac{2}{3}} e^{i \phi} \ket{1}$. 
%In this case the corresponding quantum circuit is given by Fig. 7b. In this paper we choose $\theta = \pi / 4$ and compute the $phi$-dependence of fidelity experimentally 
%and theoretically. From the results of the previous section the fidelity exhibits an oscillatory behavior. The result is summarized in Table III. 

\begin{center}
\begin{tabular}{c|c|c|c}  \hline  \hline
$\phi$ & Fidelity (qiskit Exp.) & Fidelity (ibm$\_$oslo Exp.) & Fidelity (Theory)   \\  \hline
$0.5$  &  $0.65799$  & $0.66021$  &  $0.66084$    \\
$1.0$  &  $0.63380$  & $0.63492$ &  $0.63427$    \\
$1.5$  &  $0.61851$  & $0.61904$  &   $0.61833$   \\
$2.0$  &  $0.62822$  & $0.62817$  &  $0.62768$    \\
$2.5$  &  $0.64997$  & $0.65429$  &  $0.65371$    \\
$3.0$  &  $0.67718$  & $0.67552$  &  $0.67251$    \\
$3.5$  &  $0.66590$  & $0.66869$  &  $0.66678$    \\   \hline  \hline
 \end{tabular}
\\
\end{center}

\begin{center}
{\large{Table V}}: Experimental and Theoretical Fidelities when $\ket{\psi}_A = \sqrt{1/3} \ket{0} + \sqrt{2/3} e^{i \phi} \ket{1}$ when $\theta = \pi / 4$.
\end{center}

\end{appendix}

\newpage

\begin{appendix}{\centerline{\bf Appendix C: Numerical results for Bell Measurement of $\{0,5\}$ qubits }}
\setcounter{equation}{0}
\renewcommand{\theequation}{C.\arabic{equation}}

\begin{center}
\begin{tabular}{c|c|c|c}  \hline  \hline
$\theta$ & Fidelity (qiskit Exp.) & Fidelity (ibm$\_$oslo Exp.) & Fidelity (Theory)   \\  \hline
$0.1$  &  $0.989743$  & $0.99006$   &  $0.990132$    \\
$0.2$  &  $0.961761$  & $0.96229$ &  $0.962036$    \\
$0.3$  &  $0.914808$  & $0.92463$  &  $0.919754$    \\
$0.4$  &  $0.867469$  & $0.87076$  &  $0.868720$    \\
$0.5$  &  $0.817485$  & $0.80486$  &  $0.814669$    \\
$0.6$  &  $0.759210$  & $0.76478$  &  $0.763203$    \\
$0.7$  &  $0.725925$  & $0.71541$  &  $0.720053$    \\
$0.8$  &  $0.692364$  & $0.69252$  &  $0.691416$    \\
$0.9$  &  $0.691049$  & $0.67408$  &  $0.683556$    \\
$1.0$  &  $0.701537$  & $0.69775$  &  $0.701174$    \\
$1.1$  &  $0.738694$  & $0.74922$  &  $0.744949$    \\
$1.2$  &  $0.813791$  & $0.81428$  &  $0.809426$    \\
$1.3$  &  $0.883570$  & $0.88564$  &  $0.882660$    \\
$1.4$  &  $0.950032$  & $0.94827$  &  $0.948429$    \\
$1.5$  &  $0.990324$  & $0.99016$  &  $0.990640$    \\   \hline  \hline
\end{tabular}
\end{center}
 
\begin{center}
{\large{Table VI}}: Experimental and Theoretical Fidelities when $\ket{\psi}_A = \sqrt{1/3} \ket{0} + \sqrt{2/3} \ket{1}$.
\end{center}

\vspace{0.5cm}

\begin{center}
\begin{tabular}{c|c|c|c}  \hline  \hline
$\phi$ & Fidelity (qiskit Exp.) & Fidelity (ibm$\_$oslo Exp.) & Fidelity (Theory)   \\  \hline
$0.5$  &  $0.736090$  & $0.73257$  &  $0.737904$    \\
$1.0$  &  $0.773892$  & $0.78220$  &  $0.777756$    \\
$1.5$  &  $0.807555$  & $0.79623$  &  $0.801666$    \\
$2.0$  &  $0.790106$  & $0.78980$  &  $0.787652$    \\
$2.5$  &  $0.742432$  & $0.74797$  &  $0.748597$    \\
$3.0$  &  $0.721619$  & $0.71828$  &  $0.720410$    \\
$3.5$  &  $0.732799$  & $0.72106$ &  $0.729004$    \\   \hline  \hline
\end{tabular}
\end{center}

\begin{center}
{\large{Table VII}}: Experimental and Theoretical Fidelities when $\ket{\psi}_A = \sqrt{1/3} \ket{0} + \sqrt{2/3} e^{i \phi} \ket{1}$ with  $\theta = \pi / 3$.
\end{center}

\end{appendix}

\end{document}